# Strongly Anisotropic Thermal and Electrical Conductivities of Self-assembled Silver Nanowire Network

Zhe Cheng[1,2], Meng Han[1], Pengyu Yuan[1], Shen Xu[1], Baratunde A. Cola[2], Xinwei Wang[1]*

[1]Department of Mechanical Engineering, 2010 Black Engineering Building,

Iowa State University, Ames, Iowa, 50011, USA

[2]George W. Woodruff School of Mechanical Engineering, Georgia Institute of Technology,

Atlanta, Georgia 30332, USA

*Corresponding author. Email: xwang3@iastate.edu, Tel: 515-294-2085, Fax: 515-294-3261

**Abstract**

Heat dissipation issues are the emerging challenges in the field of flexible electronics. Thermal management of flexible electronics creates a demand for flexible materials with highly anisotropic thermal conductivity, which work as heat spreaders to remove excess heat in the in-plane direction and as heat shields to protect human skin or device components under them from heating. This study proposes a self-assembled silver nanowire network with high thermal and electrical anisotropy with the potential to solve these challenges. The in-plane thermal conductivity of the network along the axial direction of silver nanowires is measured as 37 W/m-K while the cross-plane thermal conductivity is only 0.36 W/m-K. The results of measurements of electrical and thermal conductivities suggest that abundant wire-wire contacts strongly impede thermal transport. The excellent alignment of nanowires results in the same anisotropy ratio of 3 for both thermal and electrical conduction in the two in-plane directions. The ratio remains unchanged as the temperature decrease to 50 K, which indicates that wire-wire contacts lower the thermal and electrical conduction in the two directions to the same extent and their effect is independent of temperature. In addition, phonon softening markedly reduces the Debye temperatures of the network, which are fitted from the electrical resistivity data. As a result of phonon thermal conduction, the Lorenz numbers of the film in the two directions, which are approximately the same, are larger than the Sommerfeld value at room temperature and decrease as temperature decreases because of small angle scattering and the reduced phonon contribution. This nanowire network provides a solution to the emerging challenges of thermal management of flexible electronics.

## 1. Introduction

The rapid development of flexible electronics motivates the research of thermal management in these electronics.[1-3] Unlike traditional electronics built on high thermal conductivity substrates such as crystalline silicon, the electrical elements of flexible electronics are built on flexible polymer substrates with thermal conductivity two to three orders of magnitude lower. Even though the power levels of these electrical elements are much lower than those of large-scale integrated circuits, heat dissipation issues in flexible electronics, which lowers the stability and shortens the life of these electronics, are one of the emerging challenges in the thermal management of electronics.[1, 3] Moreover, flexible electronics interacts frequently with human skin which needs to be protected from the hot-spots generated by the electrical elements.[4-6] These two factors create a demand for new flexible materials with highly anisotropic thermal conductivity for thermal management, where the in-plane thermal conductivity is significantly larger than the cross-plane thermal conductivity. With these properties, the materials work as heat spreaders to remove excess heat in the in-plane direction and as heat shields to protect human skin or device components under them from heating.

A candidate of these highly anisotropic flexible materials is silver nanowire, which is flexible and non-toxic with high thermal and electrical conductivity along the axial direction.[7-11] But individual nanowires need to be scaled up as functional materials. Methods usually used to scale up silver nanowires are dip-coating, spinning or spraying coating.[10-12] The fabricated nanowire networks by these methods are randomly orientated which don't facilitate in-plane thermal transport. Large scale self-assembly of nanowires provides a solution to fabricate highly aligned nanowire networks.[13] The thermal and electrical properties of this self-assembled network are not only determined by size effect like usual nanostructures, but also determined by nanowire alignment and abundant wire-wire contacts which features its structure.

In this work we report a self-assembled silver-nanowire network for thermal management applications. To understand the physics of electrical and thermal transport in this special structure, by applying the Transient Electro-Thermal (TET) technique, we measured the thermal and electrical conduction in two in-plane directions of the network from room temperature down to 50 K, and using the Photothermal-radiation (PTR) technique and the Photoacoustic (PA) technique, we studied the cross-plane thermal conductivity at room temperature. It should be pointed out that our work presents the pioneering effort to study this aligned structure of metallic nanowire network and the effect of the abundant wire-wire contacts on thermal and electrical conduction, and this nanowire network provides solution for thermal management of flexible electronics.

**2. Sample preparation and characterization**

2.1 Sample preparation

For the preparation of silver nanowire network, silver nanowires are self-assembled by a three-phase method (oil-water-air).[13] Purchased from ACS Material, LLC, the silver nanowires are dispersed in deionized (DI) water and used as received with an average diameter of 50 nm and an average length of 200 μm. The schematic diagram of the self-assembly process is depicted in Figure 1. First, a clean beaker is half-filled with chloroform. Then, silver nanowire solutions are dropped on the chloroform surface. The amount of dropped solutions is controlled so that water phase and air phase cover approximately 80% and 20% of the chloroform surface. As shown in

Figure 1 (a-b), evaporation of the chloroform drives the water layer to move, which results in the self-assembly of silver nanowires at the surface of the water layer. The self-assembled nanowire network is a mirror-like film at the top of the water phase, which can be coated on glass slides, silicon wafers, or textiles.

After the film is assembled, chloroform is replaced by water in the fabrication of dry samples for TET measurement. Figure 1 (c-d) show that the film is coated on the beaker wall after chloroform is removed and the film is suspended again after DI water is added. Figure 1 (e) depicts the film is attached to a hard paper whose central part is cut off. After being drawn up from water, the film is dried in air for several hours. The film covering the central "window" of the paper is suspended in air, as shown in Figure 1 (f). Finally, the dry and suspended film is cut into strips for TET measurement.

2.2 Structure characterization

The structure of the self-assembled silver nanowire film is characterized by scanning electron microscopy (SEM) and x-ray diffraction (XRD). As illustrated in Figure 2 (a-b), the film is composed of two layers: one aligned layer and one randomly orientated layer. Figure 2 (b) defines directions that will be used later. Cross-plane direction is the direction through the film. Parallel and perpendicular directions are parallel and perpendicular to the axial direction of the aligned silver nanowires. Figure 2 (c-d) show the SEM graphs of the aligned and randomly orientated layers, respectively. This very aligned structure results in the anisotropic properties of the film. In addition to SEM, the nanowires were also studied by XRD. Its pattern is shown in Figure 2 (e). The crystal sizes calculated from the peaks in the pattern are 46 nm in the (111) direction, 14 nm in the (200) direction, 14 nm in the (220) direction, and 11 nm in the (311) direction. All of these sizes are smaller than the average diameter of the nanowires, which indicates that the silver nanowires are polycrystalline. The lattice plane spacing for peaks (111), (200), (220), and (311) are 2.3618 Å, 2.0314 Å, 1.4519 Å, and 1.2277 Å and the corresponding lattice constants are 4.09 Å, 4.06 Å, 4.11 Å, and 4.07 Å, respectively.

2.3 Thermal and electrical characterization

For thermal and electrical measurement, the to-be-measured sample is suspended between two electrodes (heat sinks) after being cut into strips. The TET technique [14-18] is used to measure the thermal and electrical properties of the film strip. In the measurement, a constant electrical current is applied to the sample. The joule heat generated by the current induces temperature rise of the sample. Electrical resistivity is proportional to temperature so the increase of temperature is proportional to the increase of electrical resistance of the sample and correspondingly the increase of the voltage. By measuring the voltage change, we obtain the data of temperature variation. The thermal properties of the sample are extracted by fitting the data to a theoretical thermal model. The fitted thermal conductivity is the in-plane thermal conductivity along the direction of electrical current. To measure the thermal conductivity in the two directions, in our work we prepare two strip-shaped samples. Specifically, the size of the sample, fabricated to measure the parallel thermal and electrical conductivity, is 1751 μm long, 38 μm wide, and 4.1 μm thick while the size of the perpendicular counterpart is 1456 μm long, 64 μm wide, and 4.5 μm thick. Additionally, the electrical currents applied to these samples are small (several mA) so all the temperature rises in the samples are less than 7 K. The small temperature rises guarantee that at the wire-wire contacts nanowires are not melted together during the measurement. This is also confirmed by the same electrical resistance before and after one measurement.

During the measurement process, convection and radiation also play important roles. To suppress the effect of air convection, the sample is placed in a vacuum chamber with its pressure below 0.5 mTorr. The effect of radiation is estimated as $(8\varepsilon_r \sigma T_0^3 L^2)/(D \cdot \pi^2)$, where $\varepsilon_r$ is the emissivity, $\sigma$ is the Stefan-Boltzmann constant, $T_0$ is the temperature, $L$ is the length, and $D$ is the thickness of the sample. If emissivity takes the value of 0.3, the maximum effects of radiation for the two measured samples are approximately 1% of those of thermal conduction, which are negligible in the measurement. From the thermal characterization data, we also obtain electrical resistance. To eliminate the effect of contact resistance, a four-point method is used to measure the electrical resistance. Its schematic diagram is illustrated in the top inset of Figure 3.

The in-plane thermal conductivity is characterized by the TET technique while the cross-plane thermal conductivity of the film is measured by the PTR technique.[19-21] For PTR measurement, a modulated laser is used to heat the silver nanowire film that is coated on a glass slide. The absorbed laser energy penetrates the sample through cross-plane heat conduction and emits radiation into environment. The amplitude and phase changes of the radiation signal are detected to monitor the temperature change of the sample surface. The thermal conductivity of the sample is extracted by fitting these data to a theoretical thermal model. In this work, the modulated laser beam frequencies range from 17 Hz to 20K Hz.

The cross-plane thermal conductivity was also confirmed by the PA technique.[22, 23] PA technique uses a periodically modulated laser to heat the sample surface, which generates an acoustic wave in the gas adjacent to the sample surface because of the pressure and temperature variation of the gas. A microphone detects the amplitude and phase shift of the acoustic signal. The cross-plane thermal conductivity is extracted by fitting the phase shift at a range of frequencies to a theoretical thermal model. In this work, the silver nanowire film is coated on a glass slide and 150 nm Ti is coated on the film as a transducer. Measurement frequencies range from 100 Hz to 4000 Hz. In this frequency range, the maximum penetration depths in the cross-plane and in-plane directions are 0.053 and 0.47 mm, respectively, which are smaller than the laser beam size (1 mm). Here, we consider the cross-plane heat conduction as one-dimensional conduction.

## 3. Results and discussion

3.1 Anisotropic electrical resistivity

The electrical resistivity of the film in the parallel and perpendicular directions is measured by the four-point method, and the results are shown in Figure 3. Here, we define the anisotropy ratio of electrical conductivity as $\eta_e = \sigma_{para}/\sigma_{perp} = \rho_{perp}/\rho_{para}$, where $\sigma_{para}$ and $\sigma_{perp}$ are electrical conductivity in the parallel and perpendicular directions. According to Matthiessen's rule and the Bloch-Grüneisen formula,[24] the electrical resistivity of the film is $\rho = \rho_0 + \rho_{el-ph}$. Here, $\rho_0$ is the residual electrical resistivity which is temperature independent. Its value is the resistivity value when the temperature approaches zero. $\rho_{el-ph}$ can be expressed as

$$\rho_{el-ph} = \alpha_{el-ph} \left(\frac{T}{\theta}\right)^n \int_0^{\theta/T} \frac{x^n}{(e^x-1)(1-e^{-x})} dx, \tag{1}$$

where $\alpha_{el-ph}$ is the electron-phonon coupling parameter, $\theta$ is the Debye temperature, and $n$ generally takes the value of 5 for nonmagnetic metals.[25] By fitting experimental data to the Bloch-Grüneisen formula, the Debye temperatures of the film in the parallel and perpendicular directions are 107 K and 132 K, respectively. Both are smaller than those of the bulk silver and single silver nanowire (235 K and 151 K).[7] The small Debye temperature is due to two factors: surface phonon softening in the silver nanowires, which was also observed in other nanostructures, [7, 26-30] and the abundant wire-wire contacts. The missing bonds of atoms at the defects, grain boundaries, and surfaces, lead to the change of phonon modes and the reduction in vibration frequency, which results in a reduced Debye temperature. For most of the wire-wire contacts, there exists a thin layer of organic molecular which is used to stabilize the silver nanowire suspension. Electrons need to transfer through this layer and scatter with phonons, which also contributes to a reduced Debye temperature. Additionally, Figure 3 shows that the anisotropy ratio of electrical conductivity is constant when the temperature is higher than 50 K, which indicates that temperature has the same effect on electron transport in both directions. As temperature decreases below 50 K, the anisotropy ratio increases slightly. In this temperature range, electrical resistivity, which is no longer linear with temperature, is affected by many factors. First, the effect of electron-phonon scattering diminishes while that of electron-electron scattering becomes increasingly important as temperature decreases. Second, the organic layer and thermal stress at the wire-wire contacts may result in variation of electron transport at the extremely low temperatures.

3.2. Anisotropic thermal conductivity

The TET technique is used to measure the thermal conductivity of the film in the two in-plane directions from room temperature to 50 K, and the results are shown in Figure 4 (a). Compared with the thermal conductivity of bulk silver, that of the self-assembled nanowire film in "parallel" and "perpendicular" directions decreases by 91% and 97%, respectively. Compared with the thermal conductivity of single silver nanowire, that in the two directions drop by 80% and 93%, respectively,[7] which indicates that effect of wire-wire contacts dominate thermal transport. The in-plane thermal conductivity, approximately two orders of magnitude larger than that of the flexible substrate usually used in flexible electronics, can still be enlarged significantly by reducing wire-wire contact resistance. Several possible methods include thermal annealing [31], mechanical press [32], metal coating [10], and self-limited plasmonic welding [33].

The great reduction in thermal conduction is due to the abundant wire-wire contacts and porous structure. Similarly, the high anisotropy in thermal conduction is due to the special aligned structure. In the direction parallel to the axial direction of the aligned silver nanowires, electrons transfer along the nanowires in the aligned layer while in the other direction, electrons transfer through more wire-wire contacts. Electrons transfer along the silver nanowires with smaller resistance than through the contacts. Thus, the thermal conductivity of the film shows high thermal anisotropy. Figure 4 (a) shows that the thermal anisotropy ratio is approximately constant as temperature decreases. The thermal conductivity in the "parallel" direction is about three times as large as that in the "perpendicular" direction. This ratio is the same as the electrical anisotropy ratio, which indicates that the wire-wire contacts dominantly impede thermal and electrical transport, and lower the thermal and electrical conductivity to the same extent.

As temperature decreases, the specific heat of electrons decreases and the electron mean free path increases. Because of the large number of structural imperfections such as wire-wire contacts, the increasing speed of the mean free path is smaller than the decreasing speed of specific heat. As a result, thermal conductivity of the film in the two directions decrease as temperature decreases. Specifically, the thermal conductivity of metals can be written as $\kappa = C_v v_F^2 \tau/3$, where $C_v$ is the volumetric heat capacity of electrons, $v_F$ is the Fermi velocity, and $\tau$ is the relaxation time. At low temperatures, the volumetric heat capacity of electrons changes linearly with temperature ($C_v = \gamma T$), and $\gamma$ is a constant (0.646 mJ·mol$^{-1}$·K$^{-2}$ for silver). For silver, the Fermi velocity is $1.39 \times 10^6$ m/s, and the electron density is $5.85 \times 10^{28}$ m$^{-3}$.[34] In metallic structures, both charge and energy carriers are electrons. Compared with charge transport, energy transport involves the specific heat of electrons, which is temperature dependent. Therefore, to eliminate the effect of temperature on thermal resistivity resulting from the electron heat capacity, unified thermal resistivity $\Theta = T/\kappa$ shows the effect of scattering sources on thermal resistivity, similar to electrical resistivity.[7, 15] According to Matthiessen's rule, the unified thermal resistivity can be separated into two parts based on scattering mechanisms: the phonon scattering part and the structural scattering part ( $\Theta = (3/\gamma v_F^2) \cdot (\tau_0^{-1} + \tau_{ph}^{-1}) = \Theta_0 + \Theta_{ph}$ ). As temperature approaches zero Kevin, the effect of phonon-electron scattering diminishes ( $\Theta_{ph}$=0 ). The residual unified thermal resistivity ($\Theta_0$) reflects the structural information of the sample. As shown in Figure 4 (b), the unified thermal resistivity of the film and a single silver nanowire decrease as temperature decreases. The residual unified thermal resistivity of the film in the "perpendicular" and "parallel" directions and the single silver nanowire are around 10.2, 3.6, and

0.9 mK$^2$/W, respectively, which confirms our expectation: structural disorder results in a large residual unified thermal resistivity. For a perfect bulk crystal, the residual unified thermal resistivity should be zero.

In terms of the measurement of cross-plane thermal conductivity, Figure 5 depicts the data fitting of the PTR measurement. Both the phase shift and the amplitude are used to fit for the cross-plane thermal conductivity, producing excellent fitting. The corresponding cross-plane thermal conductivity extracted from the phase shift and the amplitude are 0.35 W/m-K and 0.37 W/m-K, respectively. We take the average of these two values (0.36 W/m-K) as the cross-plane thermal conductivity of the film. These values are confirmed by the PA measurement. As shown in Figure 6, fitting between the experimental data and the theoretical model values is excellent. The thickness used in the fitting takes the value of the average thickness of the two TET samples (4.3 μm). The fitted cross-plane thermal conductivity is 0.40 W/m-K, which is 10% larger than that measured by PTR because the PA sample is coated with 150 nm Ti which enhances the cross plane thermal transport.

The cross-plane thermal conductivity and the thermal conductivity in the two in-plane directions at room temperature are plotted in Figure 7 for comparison. Specifically, the cross-plane thermal conductivity is about two orders of magnitude lower than the in-plane thermal conductivity in the "parallel" direction. This low thermal conductivity in the cross-plane direction is due to the porous and layered structure and abundant wire-wire contacts as shown in Figure 2 (b). The low thermal conductivity, combined with the high reflection surface of silver nanowire network[11], contribute to both thermal conduction and radiation insulation in the cross-plane direction.

Meanwhile, the high in-plane thermal conduction facilitate heat spreading, which is specifically applicable to thermal management of flexible electronics or works as functional coatings of professional cloth in high or low temperature environment.

3.3. Lorenz number of the silver nanowire network

The temperature dependent Lorenz numbers of the self-assembled nanowire film in the two directions are shown in Figure 8. We notice that both the electrical and thermal anisotropy ratios are about 3 and constant as temperature decreases, which indicates that wire-wire contacts dominantly impede electrical and thermal transport. The structural disorder lowers the thermal and electrical conduction to the same extent. Thus, the Lorenz numbers ($L_{Lorenz} = \kappa/(T\sigma)$) in the two directions are the same at a certain temperature. Figure 6 shows that at room temperature the Lorenz number is larger than the Sommerfeld value ($2.44 \times 10^{-8}$ $\Omega \cdot W/K^2$), which is due to phonon thermal conduction.[35, 36] Specifically, for most bulk metals, Wiedemann-Franz law holds and the Lorenz number is the Sommerfeld value because electron dominates the thermal conduction and phonon contribution is small. However, for metallic nanostructures especially the ones with large amount of disorder or imperfections, the electron mean free path is limited by the structural disorder which results in a small electron thermal conductivity. In this case the phonon contribution to thermal conductivity becomes significant because of the surface phonon modes, which results in a large Lorenz number which was also observed in other metallic nanostructures.[35-38]

As temperature decreases from room temperature down to 50 K, Lorenz number decreases because of small-angle scattering and a reduced phonon contribution to thermal transport. At low temperatures, only phonons with small wave vectors are excited so electron-phonon small angle scattering dominates in electrical and thermal transport. Small angle scattering impedes the heat transport more significantly than the charge transport, [39-41] which results in a reduced Lorenz number. Moreover, as temperature decreases, the number of the excited phonon in the nanowires and the surface phonon modes at interfaces decrease, which leads to a reduced specific heat and correspondingly a reduced phonon contribution to thermal conductivity. As a result, the Lorenz number decreases as temperature decreases.

## 4. Conclusion

This paper proposed a self-assembled silver nanowire network with high thermal and electrical anisotropy which has the potential to solve the thermal management challenges in the field of flexible electronics. The high in-plane thermal conductivity (37 W/m-K) facilitated heat dissipation while the low cross-plane thermal conductivity (0.36 W/m-K) protected human skin or device components under the film from heating. The results suggest that abundant wire-wire contacts dominantly impeded the in-plane thermal transport. For the thermal and electrical conduction in the in-plane directions, the excellent alignment of nanowires resulted in a same anisotropy ratio of 3 which was constant as temperature decreased down to 50 K, which indicated that the wire-wire contacts lowered thermal and electrical conduction in both directions to the same extent and its effect was independent of temperature. In addition, phonon softening markedly reduced the Debye temperatures of the network which were fitted from the electrical resistivity data. As a result of the phonon contribution to thermal conduction, Lorenz numbers of the film in the two directions, which were approximately the same, were larger than the Sommerfeld value. As temperature decreased, the Lorenz numbers decreased because of small angle scatterings and a reduced phonon contribution to thermal transport. This nanowire network provides solutions for thermal management of flexible electronics and functional coatings of professional cloth.

**List of figures**

Figure 1  The schematic diagram of the self-assembly process. (a-b) The self-assembly method and principle. (c) The film is coated on the wall of the glass beaker. (d) The film is suspended in DI water. (e) Hard paper is used to draw up the nanowire film. (f) The film is dried in air.

Figure 2  (a) The bilayer structure of the nanowire film. (b) The SEM graph of the film cross section and the definition of the three directions. (c-d) The SEM graphs of the aligned and random layers. (e) The XRD pattern of the film.

Figure 3  (a) The temperature dependent electrical resistivity in the parallel and perpendicular directions. The right coordinate axis is the anisotropy ratio of electrical conductivity. (b) The SEM graph of a sample suspended between two electrodes. The top inset shows the four point method for electrical characterization. The bottom inset shows the magnified SEM graph of the sample surface.

Figure 4  (a) The thermal conductivity of the self-assembled nanowire film in two directions (parallel and perpendicular to the axial direction of the aligned silver nanowire). The right coordinate axis shows the anisotropy ratio of the thermal conductivity in these two directions. (b) The unified thermal resistivity of the film in the "parallel" and "perpendicular" directions compared with that of a single silver nanowire. [7]

Figure 5  The experimental data and fittings of the PTR measurement.

Figure 6  The experimental data and fitting of the PA measurement.

Figure 7  The three dimensional thermal conductivity of the nanowire film at room temperature.

Figure 8  The temperature dependent Lorenz numbers of the nanowire film in "parallel" and "perpendicular" directions.

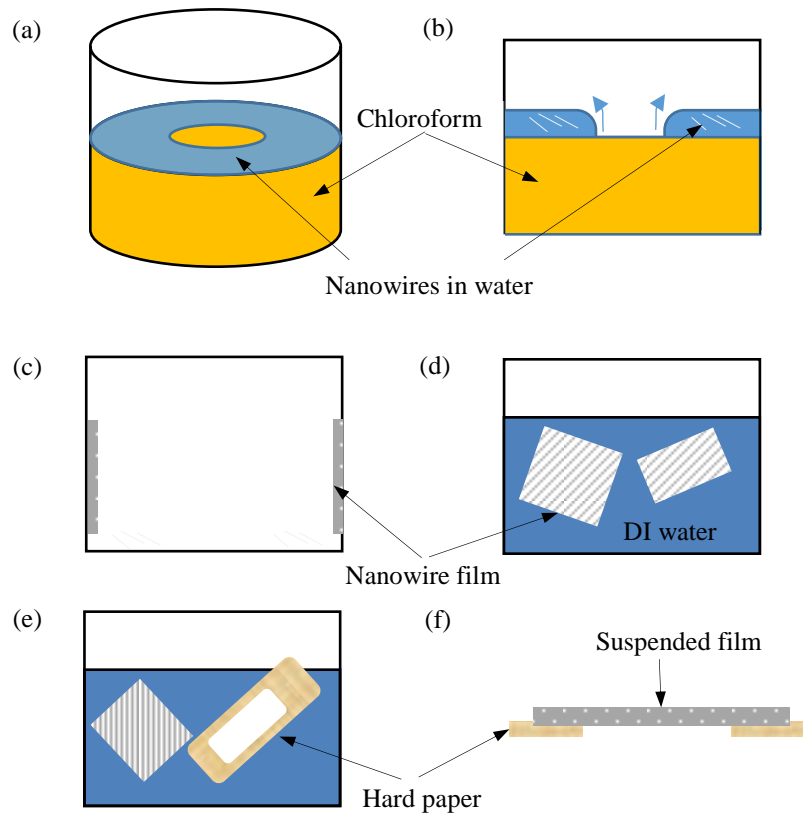

Figure 1. The schematic diagram of the self-assembly process. (a-b) The self-assembly method and principle. (c) The film is coated on the wall of the glass beaker. (d) The film is suspended in DI water. (e) Hard paper is used to draw up the nanowire film. (f) The film is dried in air.

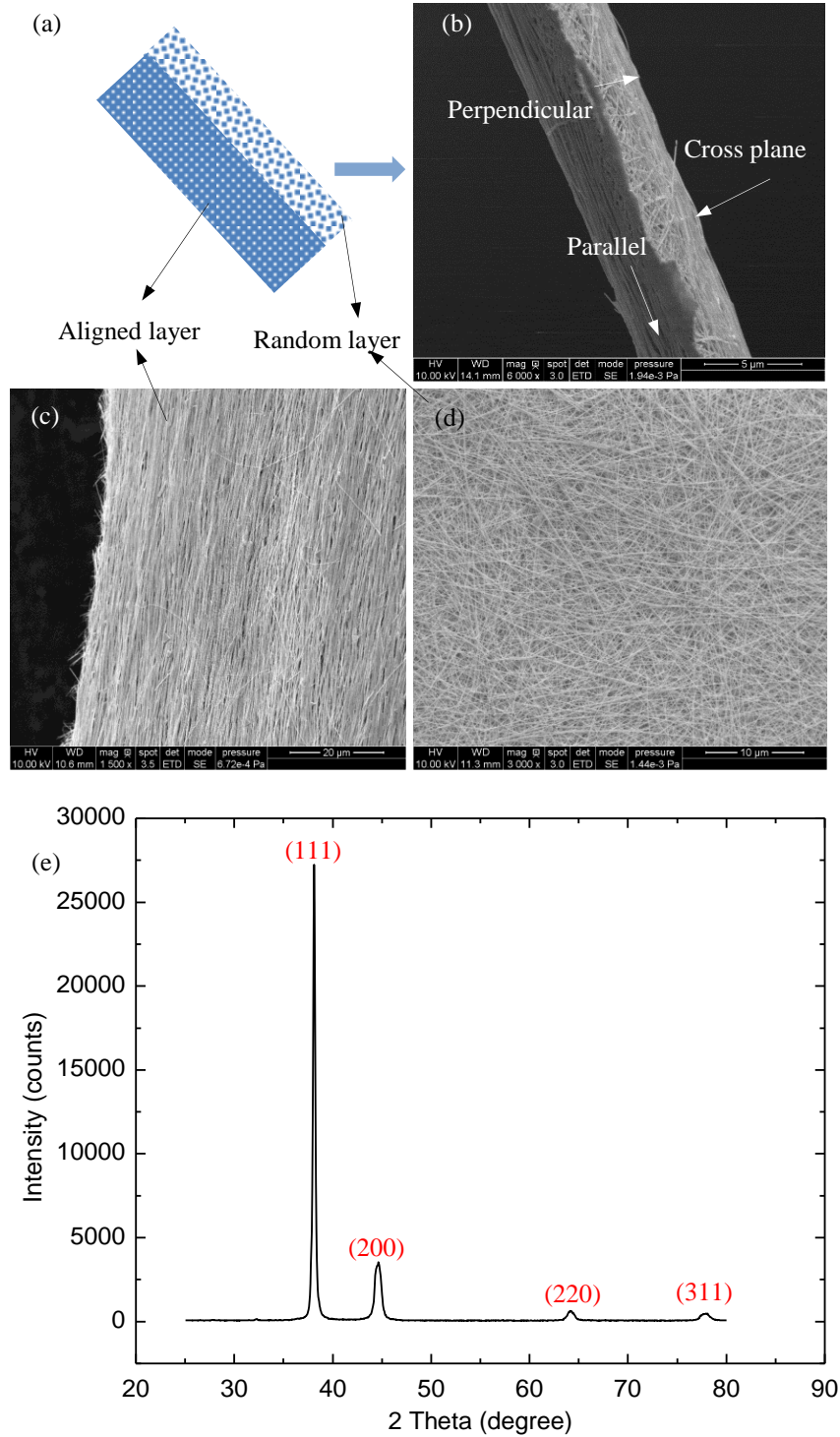

Figure 2. (a) The bilayer structure of the nanowire film. (b) The SEM graph of the film cross section and the definition of the three directions. (c-d) The SEM graphs of the aligned and random layers. (e) The XRD pattern of the film.

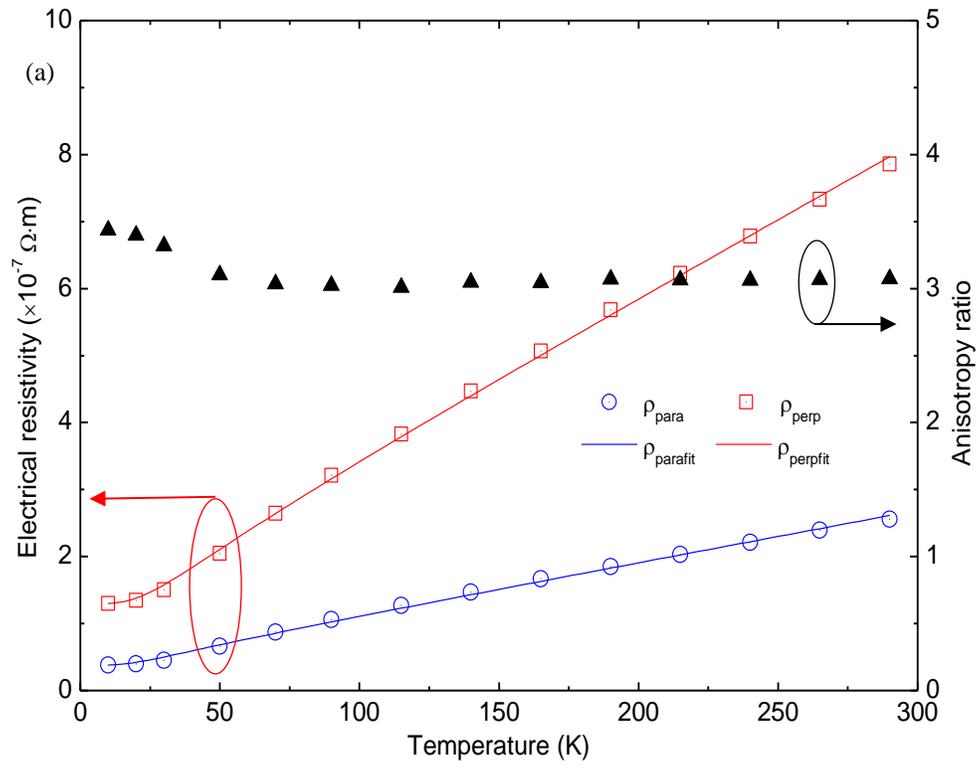

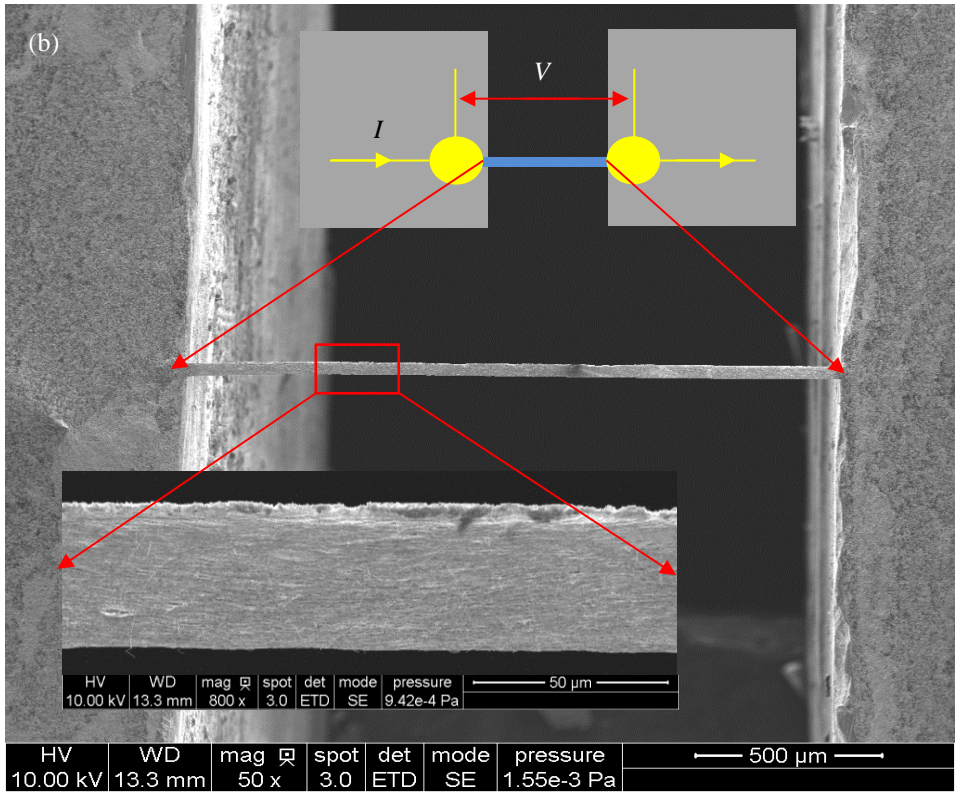

Figure 3. (a) The temperature dependent electrical resistivity in the parallel and perpendicular directions. The right coordinate axis is the anisotropy ratio of electrical conductivity. (b) The SEM graph of a sample suspended between two electrodes. The top inset shows the four point method for electrical characterization. The bottom inset shows the magnified SEM graph of the sample surface.

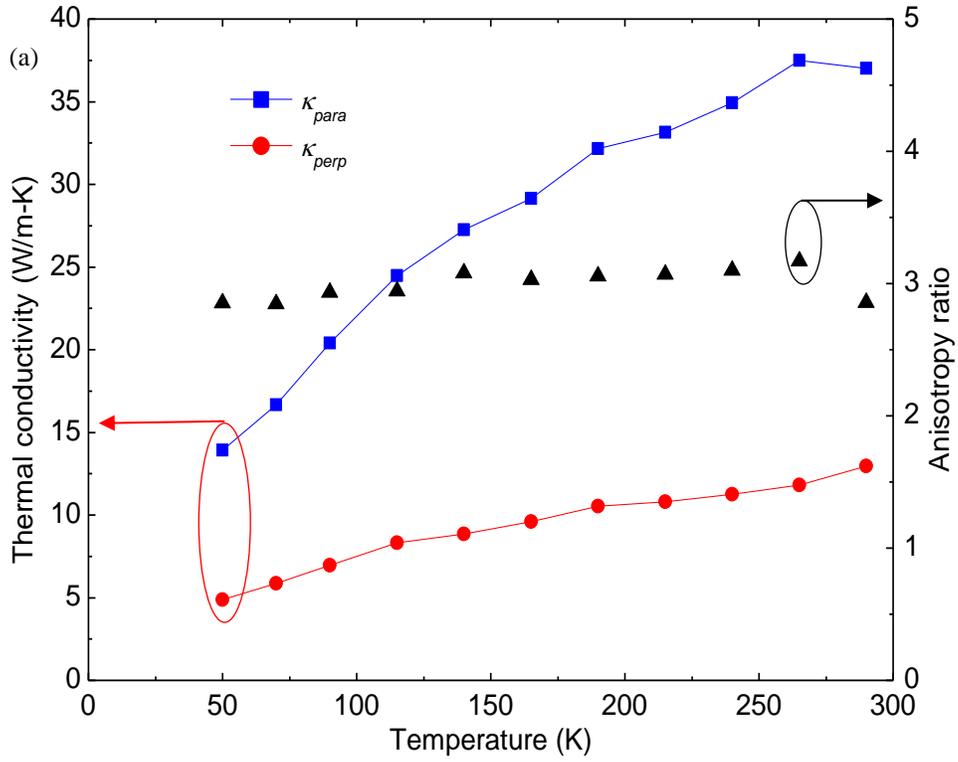

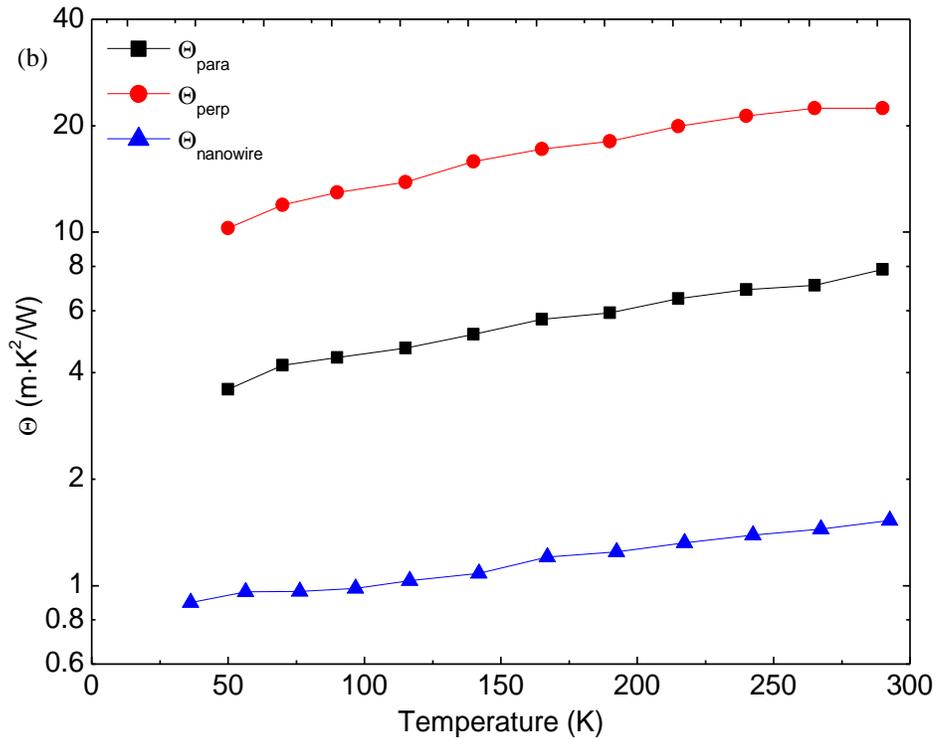

Figure 4. (a) The thermal conductivity of the self-assembled nanowire film in two directions (parallel and perpendicular to the axial direction of the aligned silver nanowire). The right coordinate axis shows the anisotropy ratio of the thermal conductivity in these two directions. (b) The unified thermal resistivity of the film in the "parallel" and "perpendicular" directions compared with that of a single silver nanowire. [7]

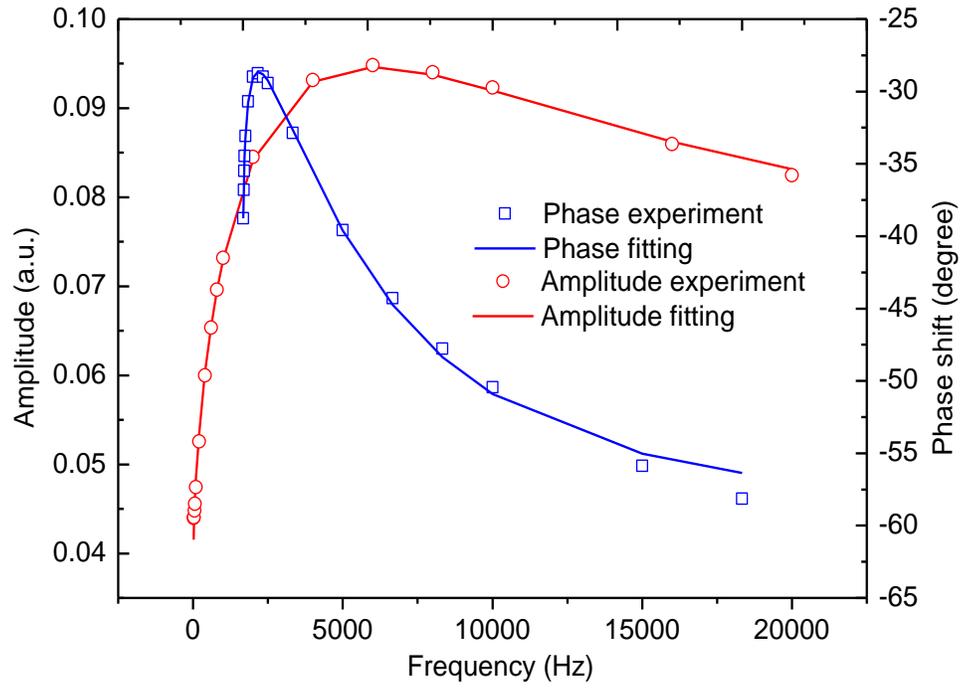

Figure 5. The experimental data and fittings of the PTR measurement.

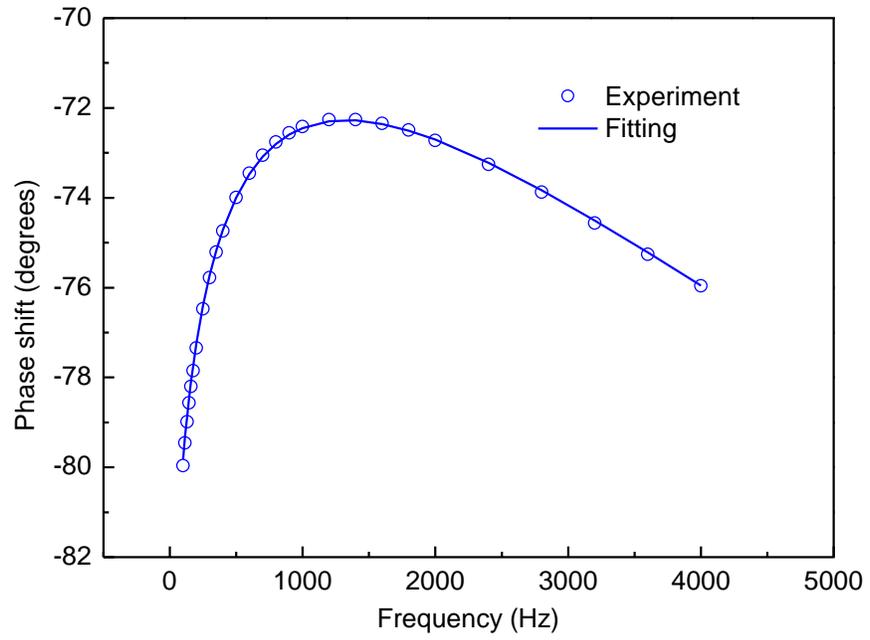

Figure 6. The experimental data and fitting of the PA measurement.

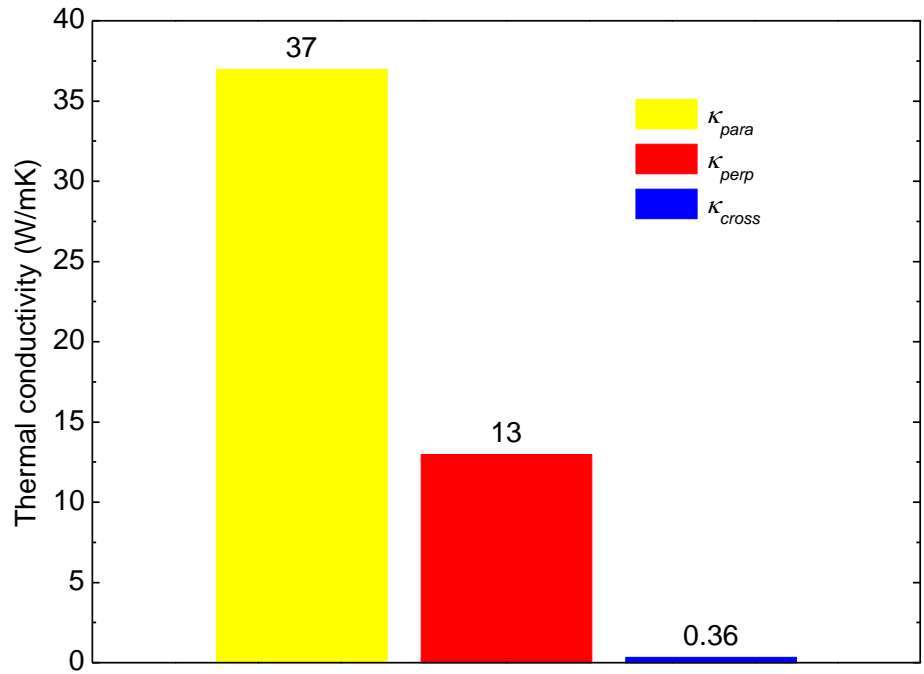

Figure 7. The three dimensional thermal conductivity of the nanowire film at room temperature.

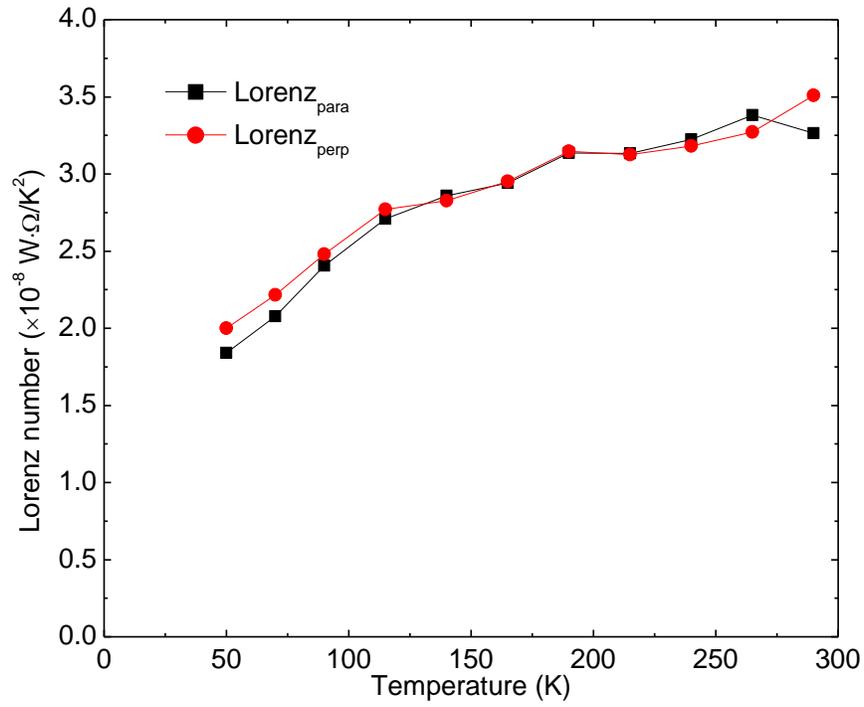

Figure 8. The temperature dependent Lorenz numbers of the nanowire film in "parallel" and "perpendicular" directions.